\documentclass[journal]{IEEEtran}
\usepackage{placeins}
\usepackage{indentfirst}
\usepackage{multirow}
\usepackage{bbding}
\usepackage{threeparttable}
\usepackage{tabularx}

\usepackage{cite}

\ifCLASSINFOpdf
  \usepackage[pdftex]{graphicx}
\else
  \usepackage[dvips]{graphicx}
\fi
\usepackage{amsmath}
\usepackage{algorithmic}

\usepackage{array}

\ifCLASSOPTIONcompsoc
  \usepackage[caption=false,font=normalsize,labelfont=sf,textfont=sf]{subfig}
\else
  \usepackage[caption=false,font=footnotesize]{subfig}
\fi

\usepackage{stfloats}
\begin{document}
%

\title{CSMCNet: Scalable Video Compressive Sensing Reconstruction with Interpretable Motion Estimation}

%
%
%

\author{Bowen Huang,
        Xiao Yan,
        Jinjia Zhou,
        and Yibo Fan
\thanks{This work was supported by Japan Science and Technology agency (JST) PRESTO Grant Number JPMJPR1757 Japan. This work was supported in part by the National Natural Science Foundation of China under Grant 62031009, in part by the Shanghai Science and Technology Committee (STCSM) under Grant 19511104300, in part by Alibaba Innovative Research (AIR) Program, in part by the Innovation Program of Shanghai Municipal Education Commission, in part by the Fudan University-CIOMP Joint Fund(FC2019-001). \emph{(Corresponding author: Yibo~Fan.)}}
\thanks{Bowen Huang, Xiao Yan and Yibo Fan are with the State Key Laboratory of ASIC and System, Fudan University, Shanghai 200433, China (e-mail: bwhuang19@fudan.edu.cn; 19112020084@fudan.edu.cn; fanyibo@fudan.edu.cn). }
\thanks{Jinjia Zhou is with the Graduate School of Science and Engineering, Hosei University, Tokyo 184-8584, Japan (e-mail: jinjia.zhou.35@hosei.ac.jp).}}

%
%

\ifCLASSOPTIONpeerreview
\else
\markboth{}%
{Huang \MakeLowercase{\textit{et al.}}: CSMCNet: Deep Unfolding for Scalable Video Compressive Sensing with Interpretable Motion Estimation Module and Interpolation Strategy}
\fi
%



\maketitle

\begin{abstract}
  Most deep network methods for compressive sensing reconstruction suffer from the black-box characteristic of DNN. In this paper, a deep neural network with interpretable motion estimation named CSMCNet is proposed. The network is able to realize high-quality reconstruction of video compressive sensing by unfolding the iterative steps of optimization based algorithms. A DNN based, multi-hypothesis motion estimation module is designed to improve the reconstruction quality, and a residual module is employed to further narrow down the gap between reconstruction results and original signal in our proposed method. Besides, we propose an interpolation module with corresponding training strategy to realize scalable CS reconstruction, which is capable of using the same model to decode various compression ratios. 
  Experiments show that a PSNR of 29.34dB can be achieved at 2\% CS ratio (compressed by 98\%), which is superior than other state-of-the-art methods. Moreover, the interpolation module is proved to be effective, with significant cost saving and acceptable performance losses.
\end{abstract}


%
\IEEEpeerreviewmaketitle

\section{Introduction}\label{sec:introduction}
%
%
%
%
\IEEEPARstart{C}{ompressive} sensing, firstly introduced by Candes, Tao and Donoho in 2006\cite{RN78}, is a novel sampling technique. This method enables exact reconstruction of original signal from much fewer samples compared with classic Nyquist sampling rule requires. Though traditional compression methods, such as JPEG and H.264, still have a wide range of application, CS has the special characteristic of sensing and compression simultaneously. This property will lead to brand new encoder and decoder systems, which has potential value in specific areas, such as single pixel camera\cite{RN79}, magnetic resonance imaging(MRI) equipment\cite{RN49}, and high-speed video system\cite{RN8}.\newline
\indent The sampling process of CS can be described as $y=\Phi x$, where $x\in R^N$ is the original signal, $\Phi\in R^{M\times N}$ is the measurement matrix, and $y\in R^M$ consists of the samples. The compression ratio can be defined as $CR=\frac{M}{N}$. In CS, fast measurement can be achieved by sampling part of $x$ with $M<<N$.\newline
\indent Theoretically, to get exact reconstruction, $x$ should have some structure-inducing regularizers\cite{RN98}, for example, the original signal should be sparse in some transformation domains\cite{RN105,RN111}, which means it can be formulated as $s=\Psi x$, where $\Psi$ is called sparsity basis or sparsity matrix, and $s$ is the sparse representation of $x$. Though it is difficult for natural images and videos to achieve true sparsity in one specific transformation domain, CS allows approximate sparsity as well, at the cost of recovery with losses.\newline
\indent With measurement matrix $\Phi$ and sparsity matrix $\Psi$, we can modify the encoding process of CS into $y=\Phi \Psi ^{-1} s$. According to CS theory, $\Phi$ and $\Psi$ should obey restricted isometry property, or RIP rule\cite{RN80,RN51}.Fortunately, RIP rule has been proved to be equivalent to sampling matrix and sparsity matrix being mutually incoherent\cite{RN97}, and random matrix is commonly chosen since it meets the requirement in most cases. Additionally, some other matrix is also used. For instance, random binary matrix is specially designed for hardware realization, and 2D-DCT matrix has more concentrated energy in the DC component\cite{RN72}. Recently, some research is done on structured random matrix(SRM), where part of random elements are replaced by manually designed structure. Experiments show that SRM can provide additional benefits, such as preserving neighbour information or reducing computation and memory consumption\cite{RN46,RN45}.\newline
\indent The CS reconstruction is to solve an under-determined linear inverse problem. Recovery methods are developed by using prior assumptions about images, such as sparsity, to narrow the solution space. The corresponding optimization model of CS reconstruction can be described as follows
\begin{equation}
    \mathop{min}\limits_{x}\;\phi(x)\;\;\;s.t.\;y=\Phi x
    \label{eq_optimization}
\end{equation}
where $\phi(x)$ is the regularization term.\\
\indent Instead of reconstruct $x$ directly, we usually solve $s$ at first, and then get $x$ by $x=\Psi^{-1} s$. The constrained standard reconstruction model of $s$ is
\begin{equation}
    \hat{s} = \arg\mathop{min}\limits_{x}{\frac{1}{2}||\Phi\Psi s - y||_{2}^{2} + \lambda ||s||_{1} }
    \label{eq_standard_recon_model}
\end{equation}
where $\hat{s}$ is the reconstruction result. Based on this model, various optimization algorithms have been exploited\cite{RN88,RN105,RN16,RN01,RN81,RN94}. Besides, driven by the powerful learning capabilities of neural networks, a number of DNN-based approaches have been applied\cite{RN54,RN8,RN270,RN35,RN90}. Classical DNN methods use deep network to map the input compressive measurements to the output reconstruction results directly. Stacked by convolution layers and activate functions, deep neural network can learn non-linear mapping from large training dataset efficiently. Moreover, algorithm unrolling\cite{RN41} is developed based on optimization methods and DNN structure. According to this technique, each iteration is mapped into a stage of the network, and multiple stages form a cascaded structure, offering promise in developing interpretable network. Several networks have applied this technique on CS reconstruction, such as ADMM\cite{RN49} and ISTA\cite{RN36}, and all achieve good experimental results in reconstruction quality and computation speed.\newline
\indent Another area of concern is that all aforementioned recovery methods are general for image and video compressive sensing, but there is additional information embedded in video sequences. Image CS reconstruction methods cannot utilize inter-frame reference of video sequences, which is widely used in traditional video compression methods. The motion estimation is important to improve reconstruction quality and our preliminary work in\cite{Huang_2020_ACCV} proposed a method that attempts to use adjacent frames to improve the reconstruction performance of video CS. To the best of our knowledge, it is the first work that utilize algorithm unrolling to develop explicit motion estimation structure in DNN for video CS reconstruction.\newline
\indent In this work, we further extend our preliminary work by adding an interpolation module,  named ITP module. This new network, dubbed CSMCNet-ITP, enables scalable reconstruction under conditions of different CS ratios with only one model. Compared with the existing deep learning based methods, CSMCNet-ITP helps saving storage in mobile devices and embedded systems, and reducing training cost. Furthermore, experimental results prove that the proposed ITP module can be added to existed reconstruction networks with only a little change, and the original recovery performance doesn't descend much. This feature adds to the versatility of our proposed method.\newline
\indent In summary, the main contributions of this paper are as follows:
\begin{enumerate}[\IEEEsetlabelwidth{3)}]
\item We use neural network modules to replace the optimization steps in the conventional model-based approaches and unfold it into hierarchical architecture that supports end-to-end training.
\item We propose a multi-hypothesis motion estimation module realized by fully-connected layers. The module generates motion estimation from reference window in the adjacent frame and helps improving the final reconstruction results.
\item An interpolation module is proposed to solve scalable sampling and reconstruction with only one model, saving memory and improving flexibility at the cost of little performance degradation. The interpolation module can be plugged into other existed deep networks, helping them to complete scalable CS tasks.
\end{enumerate}

\indent The paper is organized as follows. Section \ref{sec:background} introduces the background of block-based CS and multi-hypothesis motion estimation, and list some related works. Section \ref{sec:proposed csmcnet} and \ref{sec:proposed csmcnet-itp} describe the details of CSMCNet and CSMCNet-ITP. Section \ref{sec:experiments} shows the experimental results. In Section \ref{sec:conclusion}, we conclude this paper.
\section{Background}\label{sec:background}
\subsection{Block-based Compressive Sensing}
\indent To alleviating the huge computation and memory burdens when the input image has high resolution, block-based compressive sensing(BCS) is introduced in\cite{RN83}. According to BCS, one can break the image into smaller blocks and process each block independently. Suppose that an image is divided into $B\times B$ non-overlapping blocks, and $x_i$ is a vector representing block $i$ of input image $X$ in raster-scan fashion. The aforementioned encoding equation is modified as:
\begin{equation}
    y_{i} = \Phi_{B} x_{i}
    \label{eq_measurement_block}
\end{equation}
where $\Phi_{B}$ is a $M_{B}\times B^2$  measurement matrix and the compression rate is $CR=\frac{M_{B}}{B^2}$. It is straightforward to see that BCS in \eqref{eq_measurement_block} is equivalent to apply a whole-image measurement matrix $\Phi$ with a diagonal constraint
\begin{equation}
    \Phi = \begin{bmatrix}
        \Phi_{B} & 0 & \cdots & 0\\
        0      & \Phi_{B} & \cdots & 0\\
        \vdots & & \ddots & \vdots\\
        0 & \cdots & 0 & \Phi_{B}
    \end{bmatrix}
    \label{eq_block_phi}
\end{equation}
To coordinate with the blocked measurement matrix, the sparsity matrix $\Psi$ should also be a block-based operator of the same $B\times B$ size.\newline
\indent In this paper, the block size of 16 is used, which achieves balance of computation burdens and reconstruction performance in our method. However, BCS will produce blocking artifacts in general, and additional deblocking operations must be applied to eliminate it. Some methods\cite{RN85,RN84,RN94} use independent deblocking module, such as BM3D, and some DNN methods\cite{RN86,RN76} also perform image reconstruction and deblocking at the same time. We choose BM3D as a post-process denoiser in this article.
\subsection{Multi-hypothesis Motion Estimation}
\indent In traditional video compression methods, motion estimation(ME) and motion compensation(MC) is long employed to improve performance in H.264, H.265, etc. In consideration of bit rate limitation, these algorithms use single hypothesis MEMC in order to reduce the amount of motion vectors. However, this limitation doesn't exist in CS, since MEMC is executed at the decoder side of the system. Therefore, multi-hypothesis MEMC is considered to get better prediction results.\newline
\indent A multi-hypothesis prediction can be described as an optimal linear combination of all possible reference blocks in the search window of the reference frame
\begin{equation}
    \omega_{t,i} = \mathop{arg} \mathop{min}_{\omega} ||x_{t,i} - \tilde{x}_{t,i}(\omega)||_{2}^{2}
    \label{eq_mhme_1}
\end{equation}
\begin{equation}
    \tilde{x}_{t,i}(\omega) = \omega_{t,i} H_{t,i}
    \label{eq_mhme_2}
\end{equation}
where the subscript t is the index of frame in the video and i is the index of image block in the frame. The $H_{t,i}$ is a matrix of dimensionality $K\times B^2$, consisted of rasterizations of the $K$ possible reference blocks. In this context, $\omega_{t,i}$ is the coefficient that represents the linear combination of the possible reference blocks.\newline
\indent In the case of CS reconstruction, \eqref{eq_mhme_1} cannot be implemented directly, since $x_{t,i}$ is unknown, and only measurements $y_{t,i}$ is accessible. To solve this dilemma, \eqref{eq_mhme_1} can be modified as
\begin{equation}
    \hat{\omega}_{t,i} = \mathop{arg} \mathop{min}_{\omega} ||y_{t,i} - \Phi \omega_{t,i} H_{t,i}||
    \label{eq_mhme_3}
\end{equation}
where $\hat{\omega}_{t,i}\neq \omega_{t,i}$ generally. One commonly chosen solution to this ill-posed problem in conventional algorithms is Tikhonov regularization\cite{RN16}, while in this paper we propose a learning-based module to solve it, which appears better prediction ability.\newline
\indent Besides, some other network structure, such as RNN, is also used to extract temporal information in video CS reconstruction\cite{RN270}. With the help of algorithm unrolling, we are able to design more explicit structure to implement MEMC, and experiments show that our methods achieve superior performance due to the targeted optimization compared with the RNN based methods.
\begin{figure*}[!t]
    \centering
    \includegraphics[width=\linewidth]{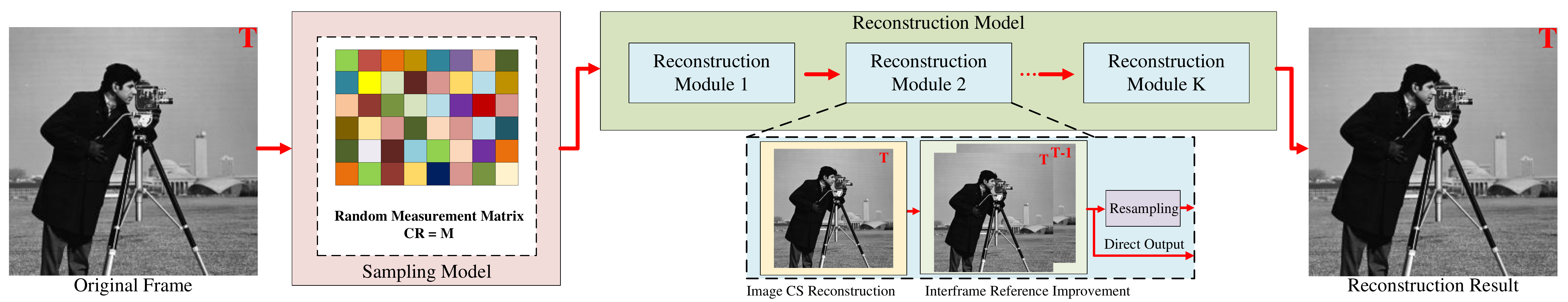}
    \caption{The sampling and reconstruction process of CSMCNet. Every reconstruction module includes image CS reconstruction and inter-frame reference improvement. Several reconstruction modules are stacked to simulate the iteration steps.}
    \label{fig_overview}
\end{figure*}
\begin{figure}[!t]
    \centering
    \includegraphics[width=\linewidth]{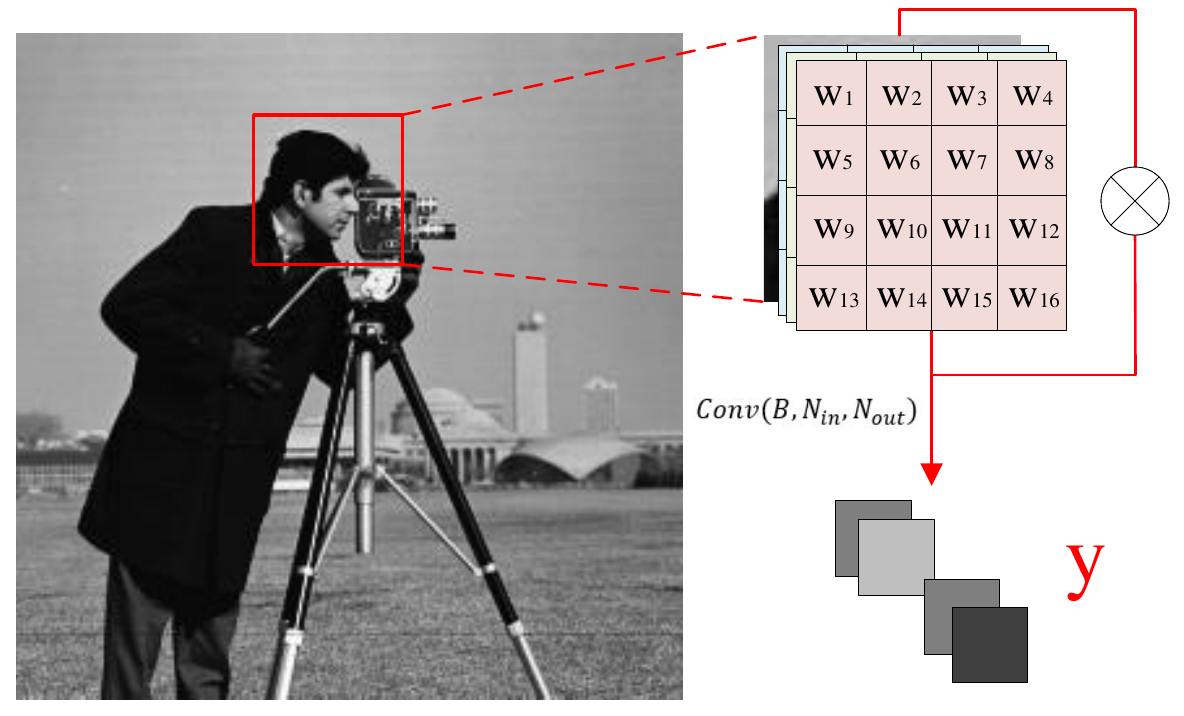}
    \caption{The measurement process of CSMCNet based on BCS. CS measurements are sampled through convolution.}
    \label{fig_sampling}
\end{figure}
\subsection{Related Works}
\indent Roughly speaking, there are three kinds of recovery algorithms that have been developed, i.e. optimization-based iterative algorithms, end-to-end DNN methods and hybrid reconstruction methods.\newline
\indent For optimization-based reconstruction methods, the general idea is to find efficient regularization terms inspired by image priors. Sparsity in some transformation domains, such as DCT\cite{RN89}, wavelet\cite{RN81} and gradient domain\cite{RN88}, has been exploited to rebuild image and video signal. For instance, Li \textit{et al}. \cite{RN88} used the total variation(TV) regularizer for improving the local smoothness. Metzler \textit{et al}. \cite{RN82} combined the BM3D denoiser with the AMP algorithm, dubbed D-AMP, to achieve better reconstruction quality. For video CS, Fowler \textit{et al}.\cite{RN16} proposed a motion compensation and block-based method with projected Landweber reconstruction algorithm, in which additional reference information was obtained to help recovery. Yang \textit{et al}.\cite{RN11} used Gaussian mixture model(GMM) to recover high-frame-rate video measurements, and the reconstruction could be computed from the probabilistic model updated online. The advantage of these methods is that they usually have theoretical guarantee. However, due to the nature of iteration, the computation time of these methods is relatively long, which makes them unsuitable for real-time application. Moreover, their performance largely relies on the hand-designed prior constraints, which are difficult to decide.\newline
\indent For classic end-to-end data-driven deep neural networks, Mousavi \textit{et al}.\cite{RN85} is the first one that apply stacked denoising auto-encoder to solve CS reconstruction problem. Lohit \textit{et al}. \cite{RN84} proposed a CNN consisted of fully connected layers and convolution layers, and a BM3D denoiser was used for deblocking in postpocessing. There are also some deep networks developed for video CS specifically. Xu \textit{et al}.\cite{RN270} proposed a network named CSVideoNet, in which a multi-rate CNN and a synthesized RNN was combined to impove the trade-off between compression rate and spatial-temporal resolution of the reconstructed videos. Iliadis \textit{et al}.\cite{RN8} designed a deep fully-connected network for video compressive sensing, and the measurement matrix was trained with the recovery network coherently to improve the final results.\newline
\indent Compared with iterative methods, these networks benefits from their feed-forward architecture, making inference of results very fast. Additionally, parameters learned through training achieve better performance. Nevertheless, suffering from the black box characteristic of these models, there is no good interpretation and theoretical guarantee for these methods, bringing difficulty in making targeted improvements. Thus, it will be beneficial to integrate the ideas of conventional algorithms and deep neural networks.\newline
\indent For scalable CS, some networks aiming at multi-rate CS competent for the task. Shi \textit{et al}.\cite{RN76} proposed a scalable CNN called SCSNet. The network was composed of several sub-networks, each of which solved one specific compression rate to provide coarse granular scalability. A greedy algorithm was used to accomplish finer granular scalability. Xu \textit{et al}.\cite{RN63} proposed a scalable Laplacian pyramid reconstructive adversarial network(LAPRAN) that enabled flexible and fast CS images reconstruction. It used a hierarchical structure of multiple stages, and each stage could generate an output of different resolution at different compression rate. However, the network structure of multi-rate CS is more complex than classic ones, leading to larger model size comparably.\newline
\indent Apart from the above two types of methods, some research combines the prior knowledge and the structure of deep networks. Deep unfolding method unrolls iterative algorithms to integrate the advantages. For example, Zhang \textit{et al}. \cite{RN36} developed a network inspired by the Iterative Shrinkage-Thresholding Algorithm(ISTA). The network cast ISTA into deep network by solving the proximal mapping associated with the sparsity-inducing regularizer using nonlinear transforms. Wen \textit{et al}. \cite{RN87} unfolded the denoising process of approximate message passing algorithm(AMP), and an integrated deblocking module and a jointly-optimized sampling matrix were used to enhance performance. Nevertheless, no video CS reconstruction algorithm is unrolled and mapped into end-to-end network, and it is also difficult for these unfolding methods to perform adaptive reconstruction tasks due to fixed network size.\newline
\indent Chang \textit{et al}.\cite{RN56} proposed a general framework to train a single deep neural network that solves arbitrary linear inverse problems, including CS recovery. It was implemented through plug-and-play and the deep network acted as a quasi-projection operator of ADMM for the set of natural images. Veen \textit{et al}.\cite{RN96} used untrained deep generative models, which was based on Deep Image Prior(DIP). According to DIP, the convolution weights of the network were optimized to match the observed measurements. This method does not require pre-training over large dataset, at the cost of losing reconstruction accuracy.
\section{Proposed CSMCNet}\label{sec:proposed csmcnet}
\indent In this section, details of CSMCNet will be presented. As shown in Fig.\ref{fig_overview}, the proposed CSMCNet is composed of an encoder(sampling model) and a decoder(reconstruction model). The sampling model is based on BCS and the reconstruction model contains several stacked stages.\newline
\indent Single channel video signal is mainly focused for simplicity in our paper and we declare that colorful signal can be processed channel by channel independently.
\subsection{Sampling Model}
\indent The measurement process is implemented block-by-block as shown in Fig.\ref{fig_sampling}. Different from our previous work\cite{Huang_2020_ACCV}, where a fully connected layer was employed, we use a convolution layer to sample frames here, since convolution can be performed on input frames of arbitrary size. It can be easily proved that these are the two equivalent forms of \eqref{eq_measurement_block}. We denote this process as
\begin{equation}
    y_{t,i} = W_s\bigotimes x_{t,i}
    \label{eq_measurement_conv}
\end{equation}
where $W_s$ corresponds to $M_B$ filters of kernel size $B\times B$, the subscript t is the index of frames in the video and i is the index of image blocks in the frame. The stride of the convolution layer is set to be $B$ to realize non-overlapping BCS sampling, and the compression ratio can be adjusted by change the number of filters. There is no bias and no activation function.\newline
\indent Research has shown that different sensing matrix, such as random Gaussian Matrix, 2D-DCT matrix and 2D-WHT matrix, has essential impact on the final results, regardless of the reconstruction algorithm\cite{RN72}. An obvious reason for this influence is the fraction of the signal energy which is captured by the measurements with each sensing matrix. The 2D-DCT measurement matrix captures a much larger fraction of the signal energy, because most of the signal energy is concentrated in the low frequencies and the random matrix with normal distribution has similar effect.\newline
\indent It is worth noting that the first measurement, or the DC component, is similar in these matrices. Therefore, we refer only to the AC component when speaking about signal energy and measurements. In some research, jointly-trained sampling model is used to further improve recovery quality\cite{RN8,RN87,RN90,RN84,RN76}. In this paper, in consideration of simplicity and flexibility, we choose random normal distribution matrix as the sampling matrix and convert it into the convolution form as described in \eqref{eq_measurement_conv}.
\begin{figure}[!t]
    \centering
    \includegraphics[width=\linewidth]{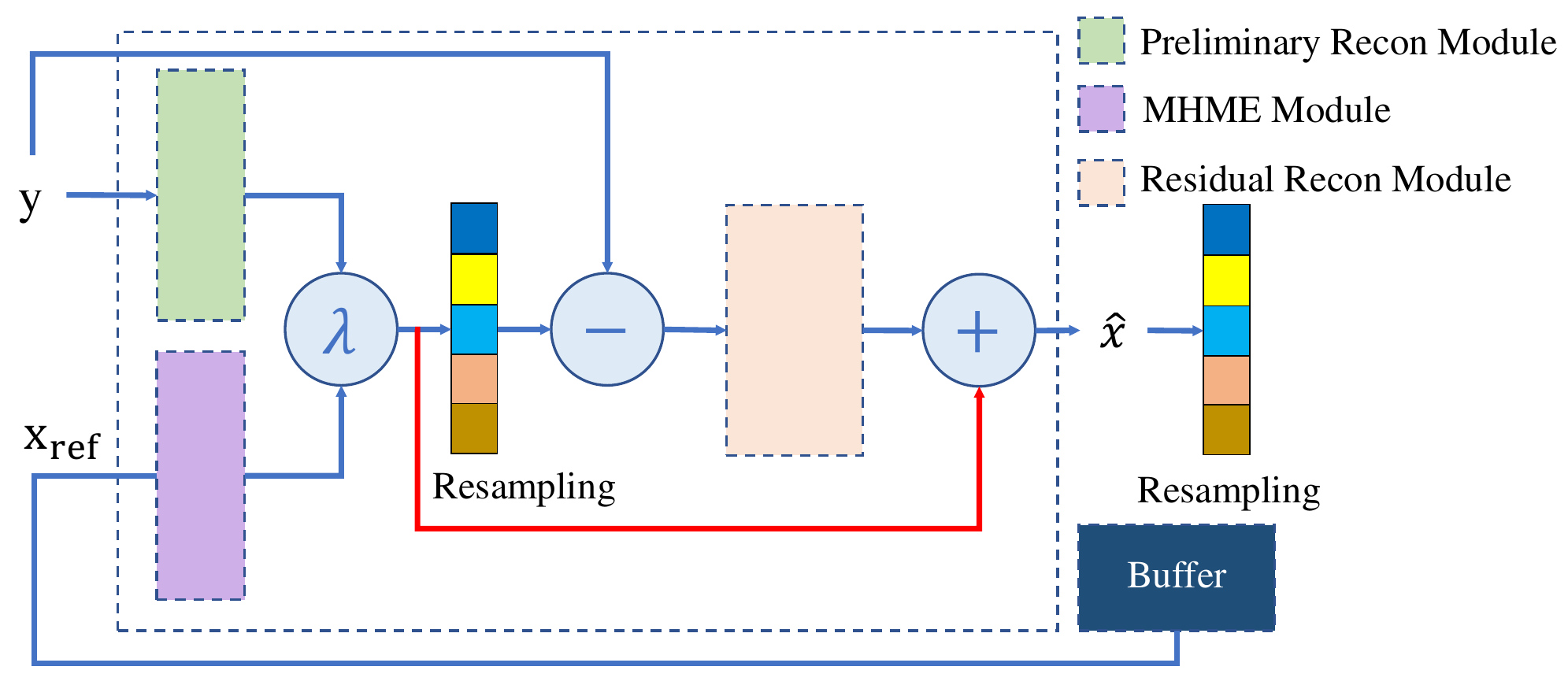}
    \caption{The structure of the i-th reconstruction stage of CSMCNet}
    \label{fig_reconstruction_stage}
\end{figure}
\subsection{Reconstruction Model}
\begin{figure}[!t]
    \centering
    \includegraphics[width=\linewidth]{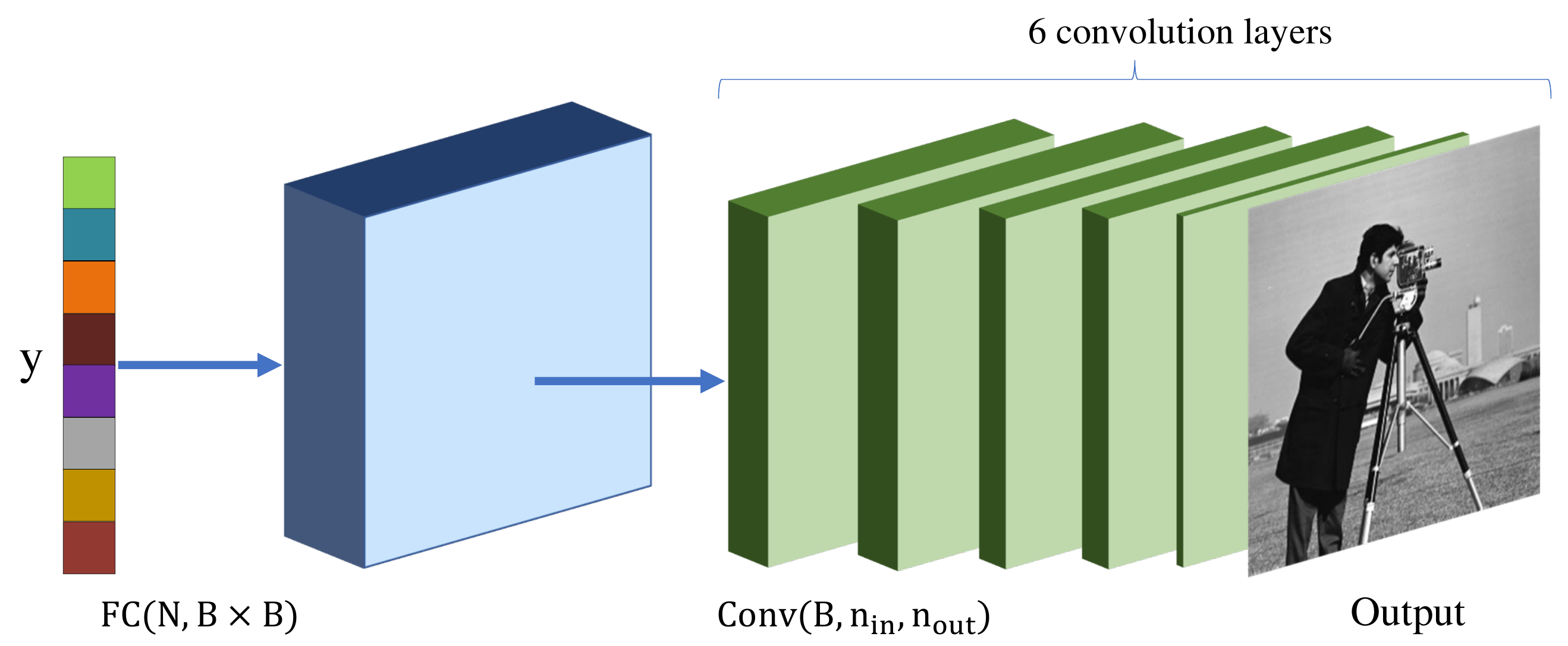}
    \caption{The structure of the preliminary reconstruction module in CSMCNet.}
    \label{fig_reconstruction_preliminary}
\end{figure}

\indent Inspired by the video CS reconstruction algorithm MC-BCS-SPL\cite{RN16}, the reconstruction model is established by unfolding the iterative steps. The model is composed of several hierarchical stages and each stage contains three modules, as presented in Fig.\ref{fig_reconstruction_stage}. The reconstruction model is divided into three parts accordingly. Firstly, the preliminary reconstruction module roughly rebuilds image blocks of size $B\times B$ from input measurements $y_{t,i}\in R^{M}$. Secondly, multi-hypothesis motion estimation(MHME) module generates prediction from a single reference frame, which was reconstructed formerly. The MHMC prediction is fused with the result of the preliminary reconstruction module as
\begin{equation}
    \hat{x}_{t,i}^{mix} = \alpha \hat{x}_{t,i}^{a} + \beta \hat{x}_{t,i}^{b}
    \label{eq_reconstruction_linear}
\end{equation}
where $\hat{x}_{t,i}^{a}$ is the preliminary recovery result and $\hat{x}_{t,i}^{b}$ is the prediction generated from the reference frame. The coefficient $\alpha$ and $\beta$ have the constraint of $\alpha + \beta=1$. Empirically, we set $\alpha=\beta=0.5$. \newline
\indent Thirdly, $\hat{x}_{t,i}^{mix}$ is remeasured with the same weight as in the sampling model and is subtracted from original measurements $y_{t,i}$ to get the residual measurements. The residual measurements are recovered and the final output of the stage is derived from adding the residual reconstruction results to $\hat{x}_{t,i}^{mix}$. We will introduce each module in the following paragraphs.\newline
\indent \textit{1) Preliminary Reconstruction Module:}The preliminary reconstruction module contains one fully-connected layer and several convolution layers, as shown in Fig.\ref{fig_reconstruction_preliminary}.\newline
\indent The fully-connected layer performs the same role as described in\cite{RN54}, and the process can be formulated as
\begin{equation}
    \hat{x}_{t,i}(y) = W \cdot y_{t,i} + b 
    \label{eq_I}
\end{equation}
where $W\in R^{B^2\times N}$ is the weight of the fully connected layer. $\hat{x}_{t,i}(y)$ is a $B\times B$ reconstructed block.\newline
\indent The following convolution layers all have a ReLU layer as activation function except for the last one. All feature maps have the same size as the output signal and the pooling layer is removed to preserve as much detail as possible. This process resembles the sparse coding stage in CS reconstruction, where a subset of dictionary atoms are combined to form an estimation to the original input signal.\newline
\indent \textit{2) MHME Module:}As introduced in \eqref{eq_mhme_2}, the process of MHME can be represented as an linear combination of all possible reference blocks in the search window of the reference frame. Thus, it is tempting to think of using a fully connected layer to accomplish it. The weight of the linear combination is learned from large dataset and can achieve superior performance compared with optimization based solutions. Due to the similarity of adjacent frames, the aggregation of motion and spatial visual features can help improve the recovery quality compared with using image CS reconstruction methods on single frame independently.\newline
\indent The reference frame is the former frame that has been reconstructed. To store the reference frame, a buffer is designed as shown in Fig.\ref{fig_reconstruction_stage}. For the sake of simplicity, we do reconstruction of current frame after the reference frame was completely reconstructed. However, the search window we use is a square with twice the size centered on the current rebuild block, and only part of the reference frame is involved. Therefore, it is possible to reduce the size of the buffer in practical application by carefully designing the reconstruction order.\newline
\indent \textit{3) Residual Reconstruction Module:}To narrow down the gap between the prediction generated by MHME module and the original signal $x_{t,i}$, we use a residual reconstruction module inspired by\cite{RN16,RN35}. The residual reconstruction module is similar to the preliminary reconstruction module, and a skip connection is added between input and output, as shown in Fig.\ref{fig_reconstruction_stage}. The skip connection helps to boost the network performance, and is proved useful to the convergence of deep neural network\cite{RN92}.
\indent The final results can be collected to form a complete frame, as described in \eqref{eq_reconstruction}, where $\kappa$ is the concatenation function that concatenates all these blocks, and $h$ and $w$ represent the numbers of blocks in row and column respectively.
\begin{equation}
    \hat{X}_{t} = \kappa\begin{pmatrix}
    \hat{x}_{t,0}(y) & \cdots & \hat{x}_{t,w-1}(y)\\
    \vdots & \ddots & \vdots\\
    \hat{x}_{t,(h-1)w}(y) & \cdots & \hat{x}_{t,hw-1}(y)\\
    \end{pmatrix}
    \label{eq_reconstruction}
\end{equation}
\begin{figure}[!t]
    \centering
    \includegraphics[width=\linewidth]{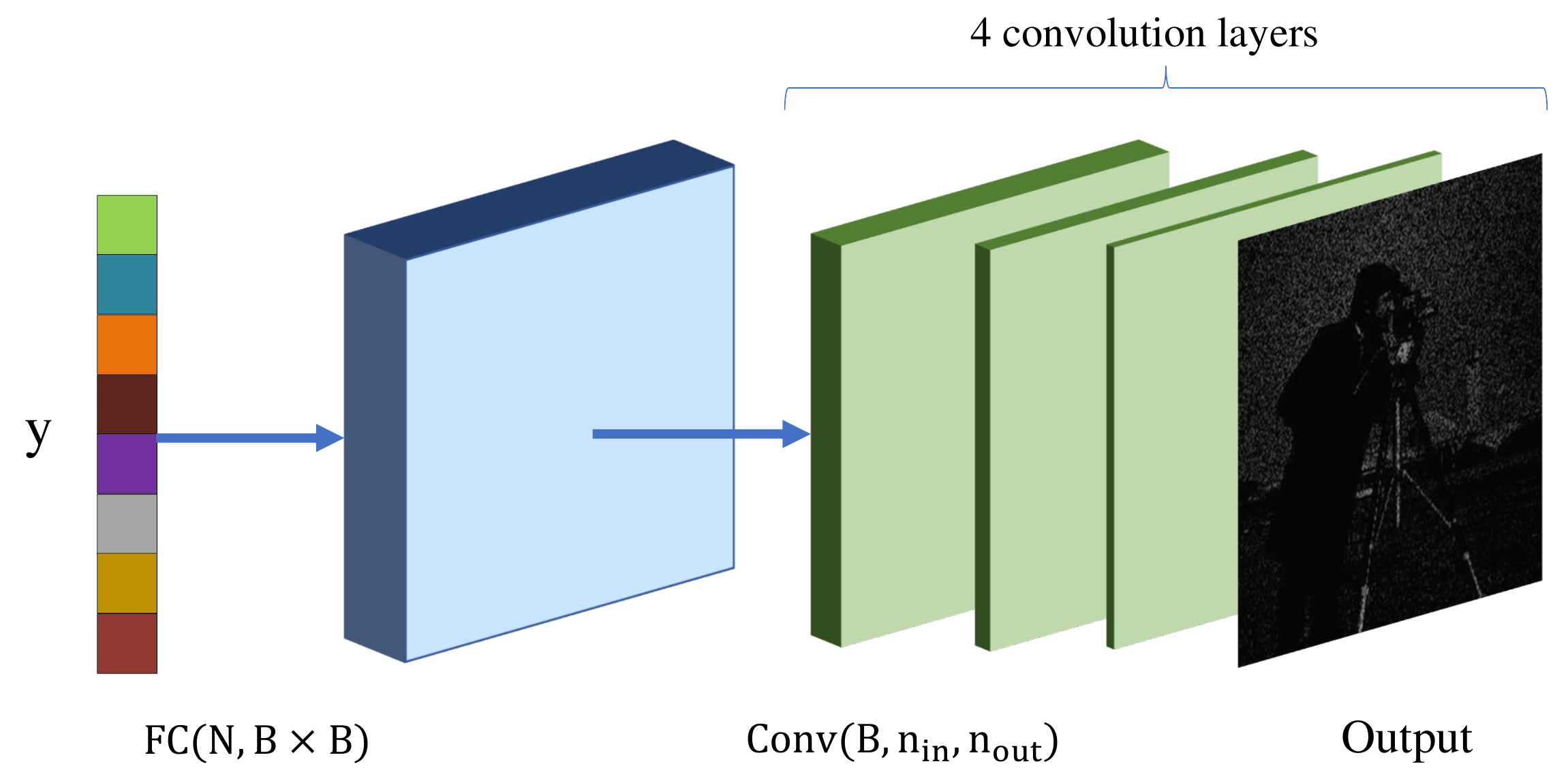}
    \caption{The structure of the residual reconstruction module in CSMCNet.}
    \label{fig_reconstruction_residual}
\end{figure}
\section{Proposed CSMCNET-ITP}\label{sec:proposed csmcnet-itp}
\indent The existing DNN methods cannot perform scalable sampling and reconstruction. To improve the flexibility of CSMCNet, an additional interpolation module, dubbed ITP module, is proposed in this section. In the interpolation module, an output selection strategy is employed and measurements of fixed size are produced. The produced measurements are sent to the following reconstruction model and experiments show that the ITP module has acceptable influence on the final results. Corresponding training strategy is used and we find that the ITP module can be easily plugged into existing networks.
\begin{figure*}[!t]
    \centering
    \includegraphics[width=\linewidth]{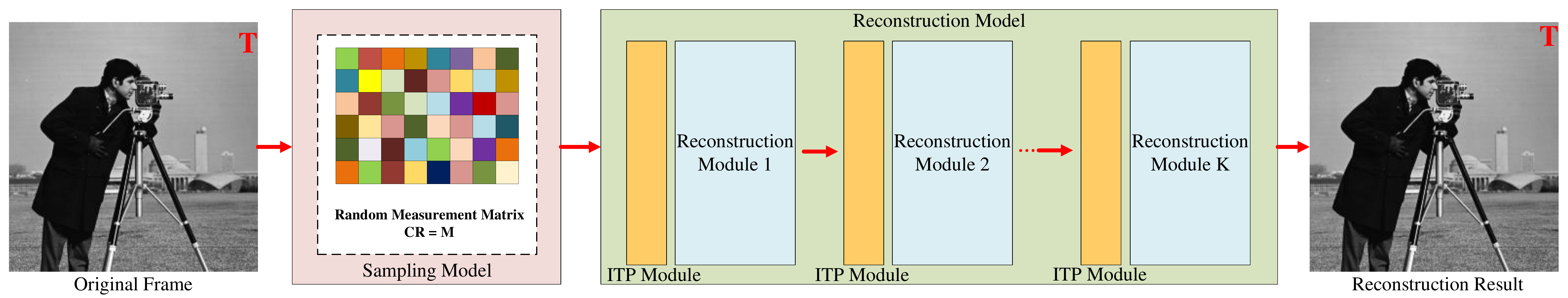}
    \caption{The illustration of ITP module in a network with several reconstruction modules. The reconstruction modules are the same as shown in Fig.\ref{fig_overview}.}
    \label{fig_itp}
\end{figure*}
\begin{figure}[!t]
    \centering
    \includegraphics[width=\linewidth]{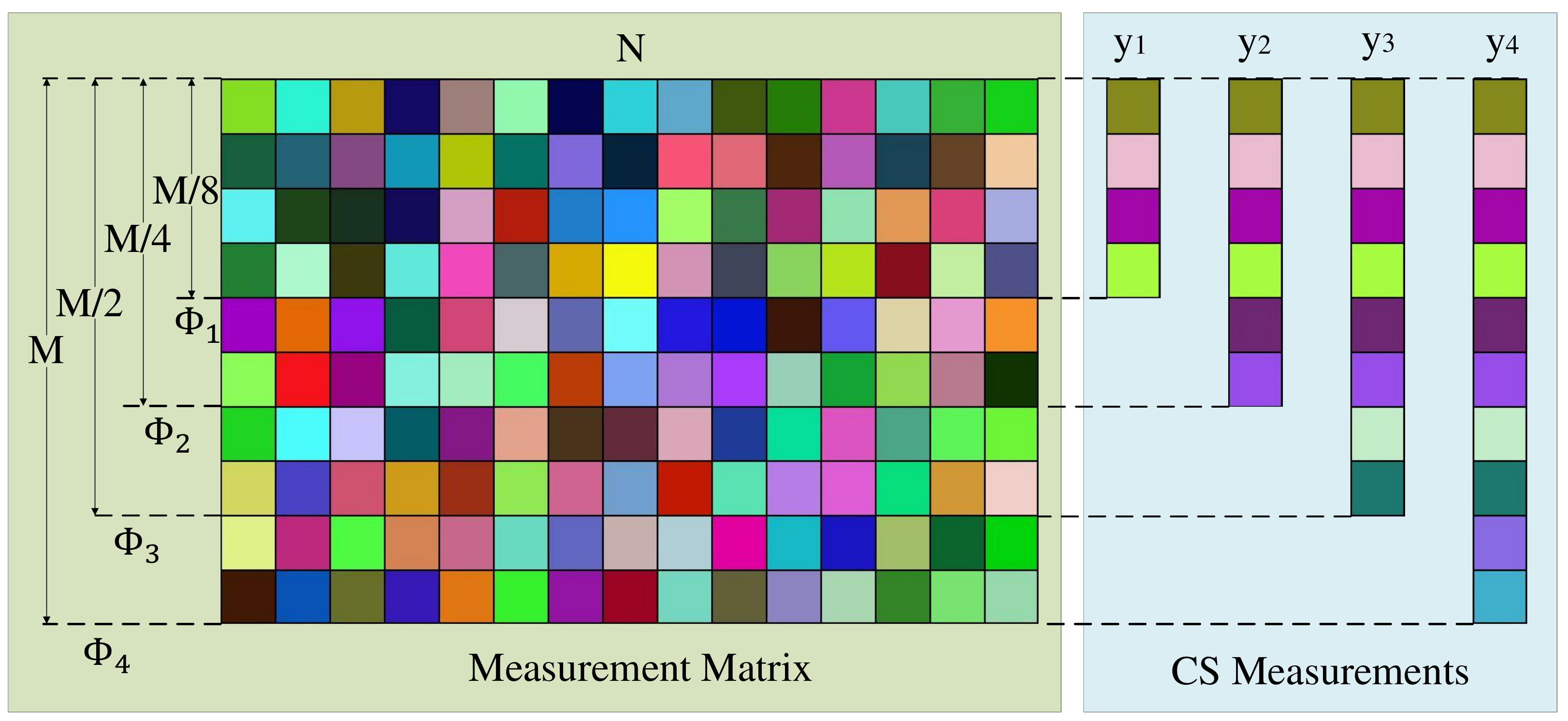}
    \caption{The illustration of a scalable sampling process with a fixed measurement matrix. The corresponding compression ratio is $66\%, 33\%, 16\%, 8\%$}
    \label{fig_sampling_random}
\end{figure}
\subsection{ITP Module}
\indent The ITP module contains one deconvolution layer and locates before the reconstruction model in the network, as illustrated in Fig.\ref{fig_itp}. Thus, the measurements obtained by the ITP module can be treated as $M_B$ feature maps. We denote the raw measurements as $y_{t,i}\in R^{M_B\times H_{C}\times W_{C}}$, where $H_{C}$ and $W_{C}$ is the height and width of the feature maps. The ITP module amplifies the number of the feature maps by a factor of $A$, which can be regarded as a 1D deconvolution. The process can be described as follows
\begin{equation}
    y_{t,i}^{\prime}=\hat{W}^{T}y_{t,i}
    \label{eq_itp_deconv1}
\end{equation}
where $\hat{W}^{T}$ is the matrix form of the weights of convolution kernel. Suppose the transpose of the weights equals to $w_{0}^{T},w_{1}^{T},...w_{n}^{T}$, then $\hat{W}^{T}$ is 
\begin{equation}
    \hat{W}^{T} = \begin{pmatrix}
    w_{0}^{T} & 0 & 0 & \cdots & 0\\
    0 & w_{1}^{T} & 0 & \cdots & 0 \\
    \vdots & 0 & \cdots & 0 & \vdots\\
    0 & \cdots & 0 & w_{n-1}^{T} & 0\\
    0 & \cdots & 0 & 0 & w_{n}^{T}\\
    \end{pmatrix}
    \label{eq_itp_deconv2}
\end{equation}
\indent We set the amplification factor $A$ to be $[\frac{CR_{max}}{CR_{min}}]$, with the predefined compression ratio ranges from $CR_{min}$ to $CR_{max}$. For $CR_{min} \leq CR < CR_{max}$, a selection strategy is employed to choose suitable number of feature maps from the interpolation results. The strategy is designed according to the feature of 1D deconvolution and can be formulated as
\begin{equation}
    y_{t,i}^{itp} = y_{t,i}^{\prime}[i \frac{CR}{CR_{min}}],\;\; i=1,2,...,n
    \label{eq_itp_selection}
\end{equation}
where the square bracket represents the index of $y_{t,i}^{\prime}$.\newline
\indent Therefore, the output of ITP module has a fixed size of $M_{max}\times H_{C}\times W_{C}$, where $M_{max}=\frac{B\times B}{CR_{min}}$. These measurements with fixed size inherit the information contained in the original measurements and can be directly used in various reconstruction models. The whole process of interpolation is fast, and the detailed information on the number of module parameters and the amount of calculations is shown in Table \ref{tab_ITP_info}. We declare that this lightweight ITP module can be easily adapted to existing DNN-based reconstruction models and provides flexibility for their deployment. Experiment results are presented in Sec.\ref{subsec:experiments3} and Sec.\ref{subsec:experiments4}.
\begin{table}[!t]
    \centering
    \caption{the detailed information of the proposed ITP module}
    \label{tab_ITP_info}
    \begin{threeparttable}
    \begin{tabularx}{\linewidth}{ l  l}
    \hline
    \textbf{ITEM} & \textbf{INFOMATION} \\
    \hline
    \hline
    Structure & 1 1D-deconvolution layer \\ 
    Input/Output Size & $L_{in}=CR\cdot B^2$ \;\;\; $L_{out}=CR_{max}\cdot B^2$\\
    Parameter Number & $K=[CR_{max}/CR_{min}]$ \\ 
    MACC* & $K\times CR\times B^2$ \\ 
    \hline
    \end{tabularx}
    \begin{tablenotes}
        \footnotesize
        \item * MACC is multiply-accumulate operations, which is commonly used as a measure of model computational complexity.
    \end{tablenotes}
\end{threeparttable}
\end{table}
\subsection{Training}
\indent A special training strategy is designed to improve the scalability of the network. We set the structure parameters of the reconstruction model according to the maximum compression ratio $CR_{max}$, and set the actual compression ratio used in the framework as a random variable, which is selected from a predefined compression ratio list. In this condition, the scalable sampling can be simply implemented by changing the number of filters of the sampling model according to the selected compression ratio, as described in Fig.\ref{fig_sampling_random}, and the ITP module will learn the interpolation coefficients efficiently.\newline
\indent One limitation of this training strategy is that it is not able to complete CS tasks of arbitrary compression ratio. The sampled signal can be correctly reconstructed only when the compression ratio $CR$ used in sampling process is chosen from the predefined list of compression ratio during training. However, we argue that sufficient $CR$ values can be added to the list during training. Five different CR value is added into the list in our experiments, and the results show that all measured signal can be rebuilded properly.
\begin{table*}[htb]
    \centering
    \caption{the detailed information of the proposed CSMCNet and other state-of-the-art methods}
    \begin{tabular}{ |>{\centering\arraybackslash}p{70 pt}|>{\centering\arraybackslash}p{50 pt} |>  {\centering\arraybackslash}p{50 pt} |>{\centering\arraybackslash}p{40 pt} |>{\centering\arraybackslash}p{60 pt}   |>{\centering\arraybackslash}p{60 pt} |>{\centering\arraybackslash}p{90 pt}|}
    \hline
    \textbf{Methods} & \textbf{Classification} & \textbf{Interpretation} & \textbf{flexibility} & \textbf{Image/Video CS} & \textbf{Joint training} &  \textbf{Other characteristic} \\ 
    \hline
    \hline
    D-AMP\cite{RN82} & Model based & \Checkmark & \Checkmark & Image CS &  & \\ 
    \hline
    MC-BCS-SPL\cite{RN16} & Model based & \Checkmark & \Checkmark & Video CS &  &motion estimation\\ 
    \hline
    ReconNet\cite{RN54} & DNN based & &  & Image CS &  & \\ 
    \hline
    ISTANet\cite{RN36} & DNN based & \Checkmark &  & Image CS &  & ISTA algorithm\\ 
    \hline
    AMPNet\cite{RN87} & DNN based & \Checkmark &  & Image CS & \Checkmark & AMP algorithm\\ 
    \hline
    CSVideoNet\cite{RN270} & DNN based &  &  & Video CS &  & RNN\\ 
    \hline
    SCSNet\cite{RN76} & DNN based &  & \Checkmark & Image CS & \Checkmark & multi-rate CS\\ 
    \hline
    LAPRAN\cite{RN63} & DNN based &  & \Checkmark & Image CS &  & multi-rate CS\\ 
    \hline
    OneNet\cite{RN56} & Hybrid & \Checkmark & \Checkmark & Image CS &  & Plug-and-Play ADMM\\ 
    \hline
    CSDIP\cite{RN96} & Hybrid &  & \Checkmark & Image CS &  & Deep-Image-Prior\\ 
    \hline
    CSMCNet & DNN based & \Checkmark &  & Video CS &  & motion estimation\\ 
    \hline
    CSMCNet-ITP & DNN based & \Checkmark & \Checkmark & Video CS &  & motion estimation/interpolation\\
    \hline
    \end{tabular}
    \label{tab_benchmark_info}
\end{table*}

\subsection{Loss Function}
\indent In this paper, for explicit reference, we use CSMCNet-K to represent CSMCNet with K stacked stages, and CSMCNet-ITP-K to represent K stages CSMCNet with the additional interpolation module.\newline
\indent The trainable parameters of CSMCNet-K contain the weight $w$ and bias $b$ of non-linear layers in the aforementioned three modules, and for CSMCNet-ITP-K, the parameters of the deconvolution layer in the interpolation module is also trainable. Given the training data pairs $(x,x_{ref})$, we want to reduce the discrepancy between the raw signal and the reconstruction results. Additionally, we also want to enhance the prediction ability of the MHME module. Therefore, the loss function for CSMCNet-K and CSMCNet-ITP-K is designed as follows
\begin{equation}
    L(w,b) = L_{err} + \lambda L_{mc}
    \label{eq_loss_1}
\end{equation}
\begin{equation}
    \left\{  
    \begin{array}{lr} 
        L_{err} = \frac{1}{2N}\Sigma_{i}^{T}||f(y_{i};W,b) - x_{i}||_{2}^{2}\;, &\\
        L_{mc} = \frac{1}{2N}\Sigma_{i}^{T}||y_{i} - \Phi_{B}x_{mc}||_{2}^{2} \;.      
    \end{array}  
    \right.
    \label{eq_loss_2}
\end{equation}
where $\lambda$ is the scale factor to control the influence of motion compensation on the total loss. Determined by experiment, we set $\lambda$ to be 0.5 during training.\newline
\indent We choose MSE to calculate loss and the proposed network can also be adapted to other loss functions. Adaptive moment estimation (Adam) optimizer with default parameters is used to optimize the network.
\section{Experiments}\label{sec:experiments}
\indent In this section, we firstly introduce the experimental settings in Sec.\ref{subsec:experiments1}, and compare our method with the-state-of-arts CS reconstruction methods in Sec.\ref{subsec:experiments2}, including conventional iterative methods and DNN based methods, both on image CS and video CS. Then in Sec.\ref{subsec:experiments5}, CSMCnet-K with $K=1,2,3,4,5$ is tested to evaluate the influence of stage number. The generality of the ITP module is validated in Sec.\ref{subsec:experiments3}, and the performance of the ITP module is evaluated in Sec.\ref{subsec:experiments4}.
\begin{table*}[htb]
\centering
\caption{the comparison results of CSMCNet-3 with other 7 methods on video dataset}
\begin{tabular}{ |>{\centering\arraybackslash}p{80 pt}|>{\centering\arraybackslash}p{60 pt} |>{\centering\arraybackslash}p{60 pt} |>{\centering\arraybackslash}p{60 pt} |>{\centering\arraybackslash}p{60 pt} |>{\centering\arraybackslash}p{60 pt}|>{\centering\arraybackslash}p{40 pt}|}
\hline
\textbf{Methods} & \textbf{$cr=20\%$} & \textbf{$cr=10\%$} & \textbf{$cr=5\%$} & \textbf{$cr=3\%$} & \textbf{$cr=2\%$} & \textbf{speed(s)}\\ 
\hline
\hline
CSVideoNet\cite{RN270} & 38.16/0.941 & \textbf{35.41}/0.903 & 32.01/0.865 & 30.13/0.823 & 28.62/0.789 & $~10^{-1}$\\ 
\hline
AMPNet\cite{RN87} & 37.44/0.952 & 32.99/0.908 & 29.04/0.847 & 27.3/0.813 & 25.67/0.781 & $~10^{-1}$\\ 
\hline
ReconNet\cite{RN54} & 37.12/0.904 & 32.11/0.878 & 28.63/0.752 & 26.62/0.631 & 25.49/0.617 & $~10^{-1}$\\ 
\hline
ISTANet\cite{RN36} & 37.93/\textbf{0.955} & 31.64/0.877 & 29.02/0.840 & 27.31/0.793 & 25.7/0.761 & $~10^{-2}$\\ 
\hline
MC-BCS-SPL\cite{RN16} & 34.25/0.891 & 30.5/0.686 & 27.61/0.506 & 25.96/0.421 & 19.42/0.246 & $~10^{3}$\\ 
\hline
DAMP\cite{RN82} & 34.2/0.758 & 30.5/0.714 & 26.54/0.673 & 20.65/0.446 & 8.69/0.064 & $~10^{1}$\\
\hline
OneNet\cite{RN56} & 34.6/0.944 & 31.72/\textbf{0.919} & 29.08/0.866 & 26.57/0.785 & 22.31/0.617 & $~10^{3}$\\ 
\hline
CSMCNet-3 & \textbf{38.34}/0.943 & 34.76/0.909 & \textbf{32.07/0.875} & \textbf{30.89/0.857} & \textbf{29.34/0.839} & $~10^{-1}$\\ 
\hline
\end{tabular}
\label{tab_compare_video}
\end{table*}

\subsection{Experimental Settings}\label{subsec:experiments1}
\indent As there is no standard dataset for video CS, we use the vimeo-90K dataset\cite{RN13} to build our own training dataset. vimeo-90K dataset contains 89800 high quality video clips from vimeo.com. We extract only the luminance component of the video and crop the central $160\times 160$ patch from each frame. We randomly choose videos from vimeo-90K dataset and get around 20,000 pairs of data for training and validation. To simulate the situation of inference, we produced reference frames of the training data pairs by using the output of image-specific CS reconstruction methods. A dataset containing 6 grey video sequences with the resolution of $1920\times 1280$ is used to perform test on video CS reconstruction, and Set11\cite{RN84} is used to perform test on image CS reconstruction.\newline
\indent Our model is realized with PyTorch and all the experiments are performed on a workstation with an Intel Xeon CPU and a Nvidia GeForce RTX2080 GPU. In the experiments, we set the block size as $16\times 16$ and choose $CR=20\%,10\%,5\%,3\%,2\%$. We normalize the input pre-feature to zero mean and standard deviation one, and all trainable parameters of the network are initialized randomly with the default initialization function of PyTorch. All models are trained for 1,000,000 iterations with batch size 250 and learning rate 0.001. In the following sections, K is set to 3 for CSMCNet-K, since CSMCNet-3 achieves the best performance as evaluated in Sec.\ref{subsec:experiments5}.\newline
\indent The iterative methods used in the comparison are D-AMP\cite{RN82} and MC-BCS-SPL\cite{RN16}; The DNN based methods are ReconNet\cite{RN54}, ISTANet\cite{RN36}, AMPNet\cite{RN87}, CSVideoNet\cite{RN270}, SCSNet\cite{RN76} and LAPRAN\cite{RN63};  Hybrid methods OneNet\cite{RN56} and CSDIP\cite{RN96} are also included. Among these approaches, MC-BCS-SPL and CSVideoNet are specially designed for video CS tasks and others are for image CS tasks. ReconNet is a classical end-to-end deep network. ISTANet and AMPNet are deep unfolding methods, and the number of layers is set to 6. SCSNet and LAPRAN are able to solve scalable CS reconstruction with one model. The summarized information about these algorithms is listed in Table \ref{tab_benchmark_info}. Two metrics, peak signal-to-noise ratio (PSNR) and structural similarity (SSIM) are used to evaluate the performance quantitatively, and the flexibility of the methods is evaluated based on time consumption and parameter number. Besides, visualized results are also presented.
\begin{table*}[!t]
\centering
\caption{the comparison results of CSMCNet-K with $K=1,2,3,4,5$ on video dataset}
\begin{tabular}{ |>{\centering\arraybackslash}p{80 pt}|>{\centering\arraybackslash}p{55 pt} |>{\centering\arraybackslash}p{55 pt} |>{\centering\arraybackslash}p{55 pt} |>{\centering\arraybackslash}p{55 pt} |>{\centering\arraybackslash}p{55 pt}|>{\centering\arraybackslash}p{65 pt}|}
\hline
\textbf{Methods} & \textbf{$cr=20\%$} & \textbf{$cr=10\%$} & \textbf{$cr=5\%$} & \textbf{$cr=3\%$} & \textbf{$cr=2\%$} & \textbf{Average}\\ 
\hline
\hline
CSMCNet-1 & 37.84/0.938 & 34.44/0.899 & 31.87/0.857 & 29.52/0.823 & 28.22/0.799 & 32.37/0.863\\ 
\hline
CSMCNet-2 & 38.27/0.942 & 34.46/0.903 & 31.92/0.866 & 30.23/0.844 & 28.24/0.821 & 32.62/0.875\\ 
\hline
CSMCNet-3 & \textbf{38.34}/\textbf{0.943} & 34.76/0.909 & 32.07/0.875 & \textbf{30.89/0.857} &29.34/0.839 & 33.08/\textbf{0.885}\\ 
\hline
CSMCNet-4 & 38.09/0.941 & 34.89/\textbf{0.911} & 32.43/\textbf{0.876} & 30.72/\textbf{0.857} & \textbf{29.55/0.840} & \textbf{33.14/0.885}\\ 
\hline
CSMCNet-5 & 38.22/\textbf{0.943} & \textbf{35.01}/0.910 & \textbf{32.49}/0.874 & 30.65/0.853 & 29.19/0.839 & 33.11/0.883\\
\hline
\end{tabular}
\label{tab_compare_stage}
\end{table*}
\subsection{Comparison with State-of-the-arts}\label{subsec:experiments2}
\begin{figure*}[htb]
    \centering
    \includegraphics[width=\linewidth]{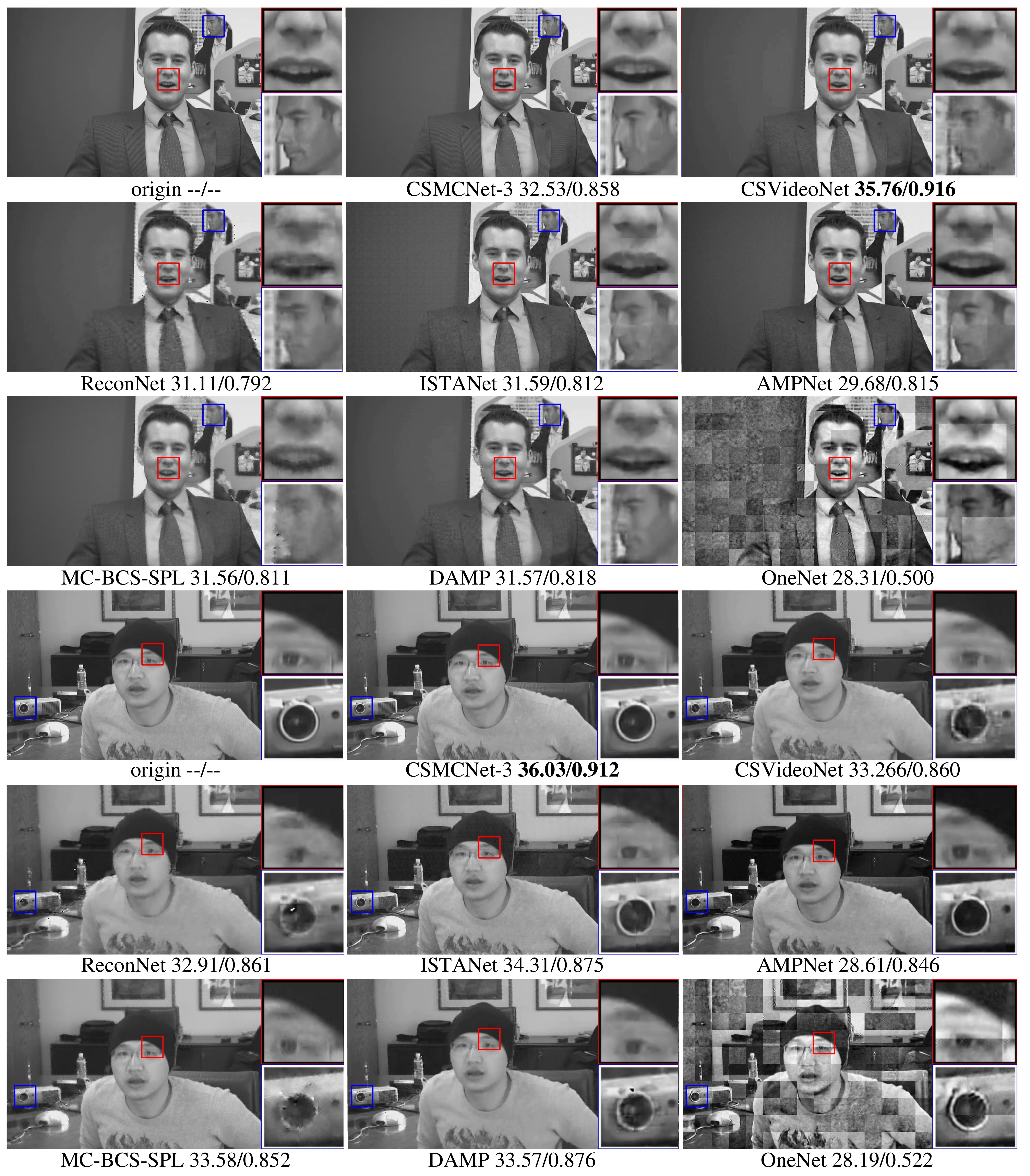}
    \caption{The visualized output of CSMCNet-3, CSVideoNet\cite{RN270}, AMPNet\cite{RN87}, ReconNet\cite{RN54}, ISTANet\cite{RN36}, MC-BCS-SPL\cite{RN16}, DAMP\cite{RN82} and OneNet\cite{RN56} on one frame of \textit{Johnny} and \textit{vidyo4} at CS ratio of $10\%$. No deblocking methods are used. PSNR/SSIM value is indicated below the figures and the best results are bolded.}
    \label{fig_compare_video}
\end{figure*}

\indent In this subsection, we compare our proposed CSMCNet-3 with other seven methods under different condition of compression ratio. The selected methods cover a variety of types, namely CSVideoNet\cite{RN270}, AMPNet\cite{RN87}, ReconNet\cite{RN54}, ISTANet\cite{RN36}, MC-BCS-SPL\cite{RN16}, DAMP\cite{RN94} and OneNet\cite{RN56}, whose characteristic can be viewed in Table \ref{tab_benchmark_info}. Before the experiments, we retrain the DNN based methods, namely CSVideoNet, AMPNet, ReconNet and ISTANet on the dataset described before and 6 video sequences with resolution of 1280x720 are used as test dataset. All the methods used in the experiments reconstruct images block-by-block and the denoising operation and sampling matrix optimization is forbidden for fairness.\newline
\indent The results of comparison are concluded in Table \ref{tab_compare_video}. We test the methods at CS ratios of $20\%,10\%, 5\%, 3\%, 2\%$. We calculate the corresponding PSNR and SSIM value as the indicator of reconstruction quality, and reconstruction speed is also considered. The best results are bolded. Since the time consumption can be affected by various factors, such as hardware performance, CPU workload, acceleration of GPU and compilation and execution speed of the programming language, we only show the order of magnitude of the speed in the table.\newline
\indent As shown in Table \ref{tab_compare_video}, generally, DNN based methods have better reconstruction quality than optimization based methods, especially for low CR conditions, verifying the effectiveness of the parameters learned by convolution layers. Besides, the speed of DNN based methods is several orders of magnitude faster than iterative methods due to their feed-forward architecture and the acceleration of GPU. Another noteworthy point is that benefited from the utilization of inter-frame information, CSMCNet-3 and CSVideoNet can further improve the recovery quality compared with other image specific DNN methods, and the performance improvement is larger at lower CS ratio. Moreover, compared with the RNN module used in CSVideoNet, the explicit MH motion estimation module employed in our model provides better reconstruction results. The reason may be that the RNN module is not purposefully optimized, and the interpretable structure has a more explicit improvement objective.\newline
\indent Visual results of the reconstruction of different methods are shown in Fig.\ref{fig_compare_video}, and PSNR and SSIM value is indicated below the images. The results are obtained under condition of $CR=10\%$, and no post denoiser is used. As presented in the visual results, the proposed CSMCNet suffers less block effect compared with other block-based CS methods, demonstrating better uniformity of our approach. Besides, our method also produces relatively richer and more accurate details in the highlight areas, leading to better subjective image quality evaluation.
\subsection{Evaluating Stages of CSMCNet-K}\label{subsec:experiments5}
\indent In this subsection, we evaluate the influence of the stage number. CSMCNet-K with $K=1,2,3,4,5$ is tested on 6 video sequences at CS ratios of $20\%,10\%,5\%,3\%,2\%$. The results are shown in Table \ref{tab_compare_stage}, and the best results are bolded.\newline
\indent As described earlier, the number of stages of CSMCNet-K corresponds to the number of iterations of the unfolded algorithm. From Table \ref{tab_compare_stage}, it can be found that as the number of stages increases from 1 to 4, the performance improves. It is understandable that deeper network gains better learning capability. However, the performance deteriorates at CSMCNet-5, and we speculate that deeper structure makes the model more difficult to converge during training. Moreover, the time consumption and memory space required to train the network also increases with more stages. Therefore, to strike a balance between the quality of reconstruction and the cost of training, we empirically decide to use CSMCNet-3 to compare with other reference methods.
\begin{figure*}[!t]
    \centering
    \includegraphics[width=\linewidth]{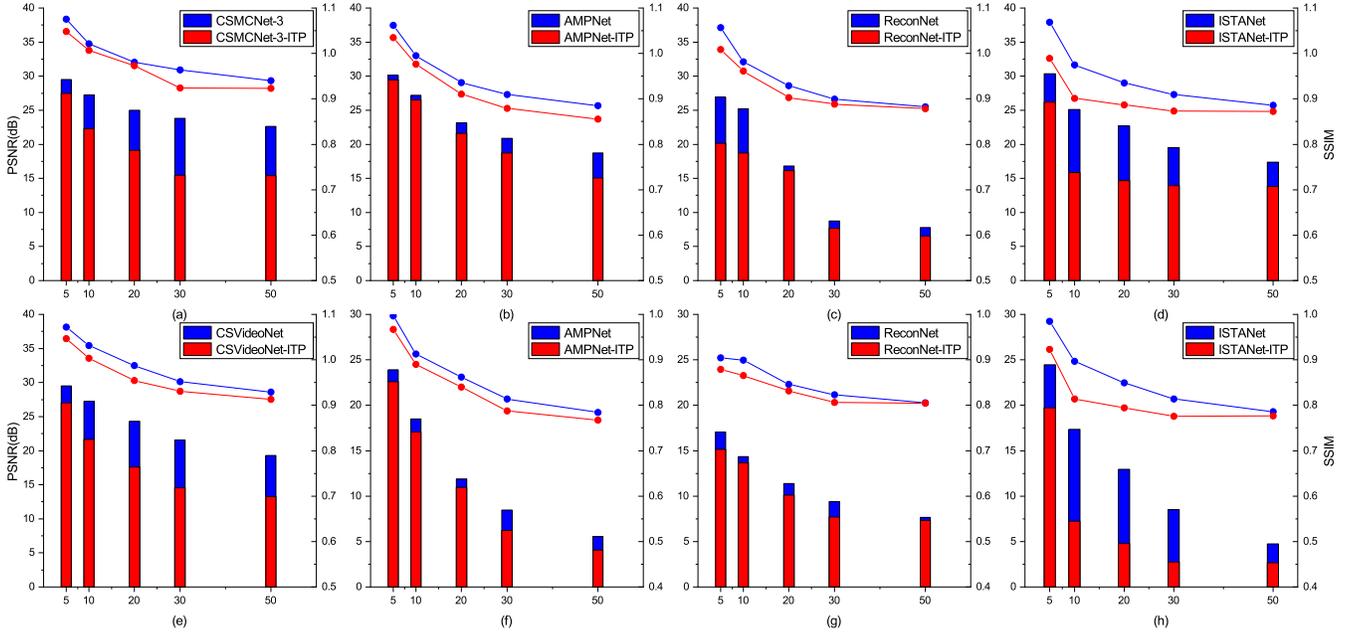}
    \caption{The results of comparison between ITP methods and original models. Sub-figures (a)-(e) are tested on video dataset and sub-figures (f)-(h) are tested on Set11. The ITP-plugged-in methods are marked in red and the original methods are marked in blue. For each sub-figure, the line+symbol graph shows the PSNR value and the bar graph shows the SSIM value.}
    \label{fig_compare_itp_unv}
\end{figure*}
\begin{table*}[ht]
\centering
\caption{the comparison results of three ITP methods with 5 other scalable CS methods on Set11}
\begin{tabular}{ |>{\centering\arraybackslash}p{80 pt}|>{\centering\arraybackslash}p{50 pt} |>{\centering\arraybackslash}p{50 pt} |>{\centering\arraybackslash}p{50 pt} |>{\centering\arraybackslash}p{50 pt} |>{\centering\arraybackslash}p{60 pt}|>{\centering\arraybackslash}p{80 pt}|}
\hline
\textbf{Methods} & \textbf{$cr=20\%$} & \textbf{$cr=10\%$} & \textbf{$cr=5\%$} & \textbf{$cr=3\%$} & \textbf{speed} & \textbf{model size}\\ 
\hline
\hline
AMPNet-ITP & 28.36/0.852 & 24.48/0.741 & 22.01/0.619 & 19.36/0.524 & $~10^{-1}$ & 3.2M\\ 
\hline
ReconNet-ITP & 23.93/0.703 & 23.26/0.673 & 21.56/0.602 & 20.33/0.554 & $~10^{-1}$ & 0.1M\\ 
\hline
ISTANet-ITP & 26.14/0.794 & 20.7/0.545 & 19.72/0.496 & 18.8/0.455 & $~10^{-2} $ & 0.5M\\ 
\hline
LAPRAN\cite{RN63} & 27.46/0.89 & 23.55/0.827 & 22.11/0.709 & 21.35/0.76 & $~10^{-1}$ & 279.9M\\ 
\hline
SCSNet\cite{RN76} & 31.82/0.921 & 28.52/0.862 & 25.85/0.784 & --/-- & $~10^{-1}$ & 4.6M\\ 
\hline
CSDIP\cite{RN96} & 23.44/0.524 & 19.53/0.391 & 17.05/0.33 & 15.17/0.166 & $~10^{2}$ & --\\ 
\hline
OneNet\cite{RN56} & 26.73/0.831 & 25.43/0.811 & 22.88/0.739 & 20.69/0.638 & $~10^{3}$ & --\\ 
\hline
D-AMP\cite{RN82} & 24.48/0.733 & 19.56/0.422 & 12.85/0.098 & 8.8/0.09 & $~10^{1}$ & --\\ 
\hline
\end{tabular}
\label{tab_compare_set11}
\end{table*}
\subsection{Validating the Universality of the ITP Module}\label{subsec:experiments3}
\indent In this subsection, we validate the generality of the ITP module. To prove that the ITP module can be easily applied to different deep learning models, we plug the ITP module into several methods, including CSMCNet-3, CSVideoNet\cite{RN270}, ReconNet\cite{RN54}, ISTANet\cite{RN36} and AMPNet\cite{RN87}. We compare their performance with the original models on both video sequences and image dataset at CS ratios of $20\%,10\%, 5\%, 3\%, 2\%$. The video sequences are the same as used in Sec.\ref{subsec:experiments2}, and the image dataset is Set11\cite{RN84}. No denoising strategy and sampling matrix optimization is used. All models without ITP module are trained seperately at different CS ratios and those having ITP module are only trained once.\newline
\indent Fig.\ref{fig_compare_itp_unv} shows the comparison results. It can be found that compared with the original methods, the plugged-in methods have relatively worse performance, and the gap is larger at higher CS ratio. The reason may be that the ITP module cannot fully rebuild the distribution of the signal via interpolation, and as the number of interpolated elements increases at low CS ratio, the gap between them and the original distribution narrows down. This phenomenon proves the validity of the interpolation strategy.\newline
\indent Though the loss of reconstruction performance is to be expected, taking consideration of the parameter size of the models and the cost of training, the loss of performance is acceptable. As the ITP module is light weighted compared to the main body of reconstruction models, we can suppose that the parameter size and time consumption of training decreases by a factor of $N$ under the same operating conditions, where the factor $N$ is the number of CS ratios we want to apply. In our experiments, the ITP method saves 5 times the storage space and training time approximately, at the cost of a $10.92\%$ drop in PSNR performance and a $8.27\%$ drop in SSIM performance in average. Overall, we think this is an acceptable price to pay.\newline
\indent Besides, Fig.\ref{fig_compare_itp_unv} also shows that for some methods, such as AMPNet, ITP module is more effective, and for others, such as ISTANet, the strategy performs poorly. Our future research will focus on which types of methods will work better with the interpolation strategy.
\begin{figure*}[!t]
    \centering
    \includegraphics[width=\linewidth]{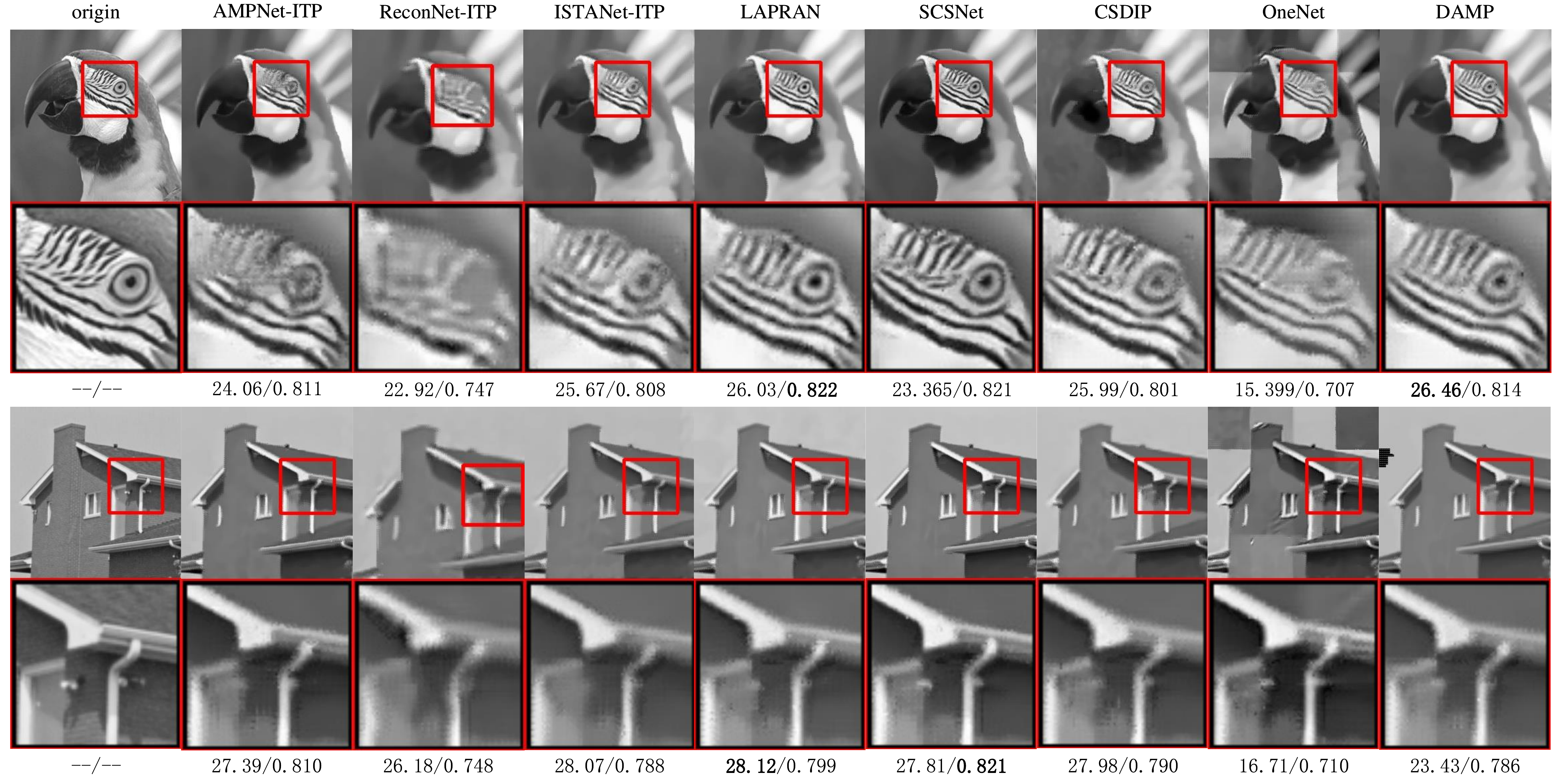}
    \caption{The visualized output of 8 adaptive CS methods on \textit{Parrots} image and \textit{house} image at CS ratios of $20\%$. The output is denoised by BM3D to alleviate block effect. PSNR and SSIM value are indicated below the figures and the best results are bolded.}
    \label{fig_compare_set11}
\end{figure*}
\subsection{Evaluating the ITP Module}\label{subsec:experiments4}
\indent In this subsection, we further compare the ITP methods with other scalable compression ratio methods, namely D-AMP\cite{RN82}, LAPRAN\cite{RN63}, SCSNet\cite{RN76}, OneNet\cite{RN56} and CSDIP\cite{RN96}. The first of them is an optimization based iterative method, the second and the third are scalable deep learning models with fixed network architecture, and the last two combines deep learning models with iterative steps. The detailed characteristic of these methods can be found in Table \ref{tab_benchmark_info}. The test dataset is Set11\cite{RN84}, and the test condition is $CS ratio = 20\%,10\%,5\%,3\%$. No sampling matrix optimization and deblocking strategy is used. In addition to PSNR and SSIM, which server as the objective image quality indicators, we also list the order of magnitude of speed for all methods and the model size for DNN based methods, which can be used as a metric to determine the ease of model deployment. The visualized results can be found in Fig.\ref{fig_compare_set11}.\newline
\indent As concluded in Table \ref{tab_compare_set11}, the ITP methods are at the forefront among all selected methods under different CR conditions, and the model size is well controlled. For D-AMP, its performance is good at high CR, but declines rapidly at low CR. The SCSNet has the best reconstruction quality, but its adaptive capability is achieved by combining sub-models dealing with different compression rates into one large model, leading to larger parameter size compared with ITP methods. The LAPRAN also achieves good performance, but the output of LAPRAN is of different size decided by the factor of compression ratio. The PSNR and SSIM results are calculated by interpolating the output of LAPRAN with bicubic algorithm, and this property is inconvenient for practical application. For OneNet and CSDIP, the disadvantage is that they are time-consuming. Besides, the visualized results of OneNet suffers severe blocking effect. We also want to point out that, as mentioned before, the performance of ITP-plugged-in methods is highly correlated with the original methods, and that is why AMPNet-ITP and ISTANet-ITP get relatively better results compared with ReconNet-ITP.\newline
\section{Conclusion}\label{sec:conclusion}
\indent In this paper, we design a deep neural network named CSMCNet based on the deep unfolding technique. The network performs video CS reconstruction and improves reconstruction quality with motion estimation. Experimental results prove the effectiveness of the proposed multi-hypothesis motion estimation module in utilizing temporal information of adjacent frames in video sequences. Moreover, an interpolation module and corresponding training strategy is designed to implement scalable sampling and reconstruction, which is useful for flexible deployment. Finally, comprehensive comparison is conducted between proposed CSMCNet-3 and state-of-the-art reconstruction methods. CSMCNet-3 outperforms the reference methods in terms of reconstruction quality and speed. Additional experiments also demonstrate the ITP module's versatility. One potential direction of our future work is to find more suitable interpolation algorithm for the ITP module, and another direction is to evaluate the relationship of the ITP module and the following reconstruction networks.


%





\ifCLASSOPTIONcaptionsoff
  \newpage
\fi



\bibliographystyle{IEEEtran}
\bibliography{bibfile.bib}
\end{document}